\documentclass[aps,preprint,showpacs, showkeys]{revtex4}%
\usepackage{amsfonts}
\usepackage{amsmath}
\usepackage{amssymb}
\usepackage{graphicx}%
\providecommand{\U}[1]{\protect \rule{.1in}{.1in}}

\begin{document}
\title[ ]{Isomorphic Hilbert spaces associated with different Complex Contours of the
$\mathcal{PT}$-Symmetric $\left(  -x^{4}\right)  $ Theory }
\author{Abouzeid M. Shalaby}
\email{amshalab@mans.edu.eg}
\affiliation{Physics Department, Faculty of Science, Mansoura University, Egypt}

\begin{abstract}
In this work, we stress the existence of isomorphisms which map complex
contours from the upper half to contours in the lower half of the complex
plane. The metric operator is found to depend on the chosen contour but the
maps connecting different contours are norm-preserving. To elucidate these
features, we parametrized the contour $z=-2i\sqrt{1+ix}$ considered in
Phys.Rev.D73:085002 (2006) for the study of wrong sign $x^{4}$ theory. For the
parametrized contour of the form $z=a\sqrt{b+i c x}$ , we found that there
exists an equivalent Hermitian Hamiltonian provided that $a^{2} c$ is taken to
be real. The equivalent Hamiltonian is $b$-independent but the metric operator
is found to depend on all the parameters $a$, $b$ and $c$. Different values of
these parameters generate different metric operators which define different
Hilbert spaces . All these Hilbert spaces are isomorphic to each other even
for parameters values that define contours with ends in two adjacent wedges.
As an example, we showed that the transition amplitudes associated with the
contour $z=-2i\sqrt{1+ix}$ are exactly the same as those calculated using the
contour $z=\sqrt{1+ix}$, which is not $\mathcal{PT}$-Symmetric and has ends in
two adjacent wedges in the complex plane.

\end{abstract}
\maketitle

\begin{figure}[ptb]
\centering
\includegraphics[width=0.5\textwidth] {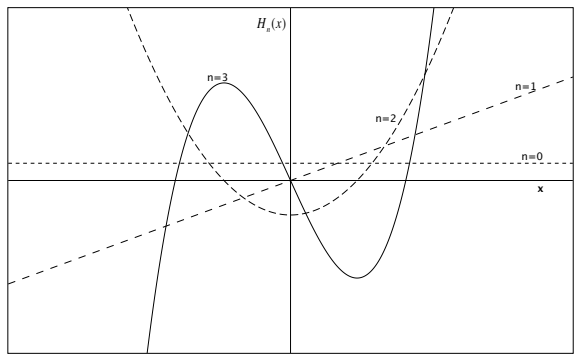} \caption{The first four
Hermite polynomials $H_{n}(x)$, $n=0,1,2,3$. One can realize that non of these
eigen functions verufy the condition $H_{n}(x)\rightarrow0$ as $\left \vert
{x}\right \vert \rightarrow \infty$.}%
\label{Hnx}%
\end{figure}

\begin{figure}[ptb]
\centering
\includegraphics[width=0.5\textwidth] {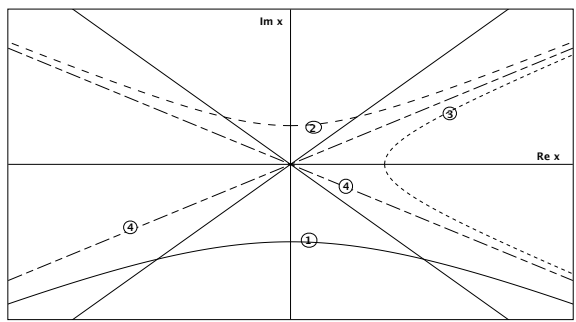} \caption{\textit{In
the lower half in this figure we have the contour $-2i\sqrt{1+ix}$ labeled by
$1$. Also, we have one of the possible roots that defines the contour
$\sqrt{ix}$ labeled by $4$ while the stokes lines are represented by solid
straight lines. Also, we have half of the contour $\sqrt{1+ix}$ labeled by $3$
and its other half lies in the upper half of the complex plane. In the upper
half, we have the contour $i\sqrt{1+ix}$ labeled by $2$, one of the possible
roots that defines the contour $\sqrt{ix}$ (dashed-straight lines) and the
other half of the contour $\sqrt{1+ix}$ . Except for the contour $\sqrt{1+ix}%
$, all the contours are $\mathcal{PT}$-symmetric and connects two non-adjacent
wedges. The contour $\sqrt{1+ix}$ is not $\mathcal{PT}$-symmetric and connects
two adjacent wedges which mean that non-of the solutions is finite at both
ends of the contour. }}%
\label{inst}%
\end{figure}

The topic of $\mathcal{PT}$-Symmetric theories is believed to solve existing
problems in Physics. This topic represents an active research area that
addresses different research directions \cite{aboebt,Abo-Lee,
aboeff,ghost1,ghost2,Symanzik,bendf,Frieder,Novel,bendmet,swanson,jonesqds,PT0,
PT2, PT3, non-norm,susy, vac}. The main stream of research in this area relies
on the fact that there exists a huge number of non-Hermitian theories which
have real spectra and thus they might have physical applications. Out of these
theories, the wrong sign $( -x^{4})$ theory is playing a vital role where its
field theoretic $( -\phi^{4})$ version represents a prototype example of a one
component scalar field theory that possesses the asymptotic freedom property
\cite{aboebt,Symanzik,bendf,Frieder}. This theory has been investigated before
and its metric operator is known in a closed form \cite{Jones}. It has been
found that the theory has an equivalent Hermitian Hamiltonian with a bounded
from below potential. In studying this theory and in all the other studies in
the literature of any $\mathcal{PT}$-Symmetric theory, a complex contour is to
be chosen within what is called the Stokes wedges of the theory \cite{Jones,
PT1}. For instance, to obtain the Hermitian form of the $\mathcal{PT}%
$-Symmetric $(-x^{4})$ theory, Jones and Mateo have employed the complex
contour $z=-2i\sqrt{1+ix}$ ( Fig.\ref{inst}). This contour has been chosen so
that it does exist in the lower half of the complex plane and starts up in a
wedge and ends in a non-adjacent one that is $\mathcal{PT}$-Symmetric to the
first. It is well known that Schrodinger equation has two independent
solutions but both of them can not decay to zero as $\left \vert x\right \vert
\rightarrow \infty$ in two adjacent Stokes wedges. However, in this work, we
use simple analysis to show that one can work with contours which are neither
$\mathcal{PT}$-Symmetric nor lie within non-adjacent Stokes wedges.
Nevertheless, we show that these contours keep all the transition amplitudes
the same as the ones obtained by using the contour $z=-2i\sqrt{1+ix}$ used in
Ref.\cite{Jones} which connects two non-adjacent wedges that are
$\mathcal{PT}$-Symmetric to each other. The idea we rely on comes from the
theory of orthogonal polynomials where the Hermite functions violate the
condition $H_{n}\left(  x\right)  \rightarrow0 $ as $\left \vert x\right \vert
\rightarrow \infty$. To illustrate this point, we consider the differential
operator in the Hermite differential equation of the form;%

\[
-\frac{d^{2}\psi}{dx^{2}}+\left(  2ixp\right)  \psi=2\lambda \psi,
\]
where $p=-i\partial/\partial x$. One can introduce the non-Hermitian
Hamiltonian $H$ such that;%
\begin{align*}
H\psi &  =2\lambda \psi,\\
H  &  =p^{2}+2ixp.
\end{align*}
Clearly, the eign functions $\psi_{n}$ are the famous Hermite Polynomials
$H_{n}\left(  x\right)  $ which do not vanish at infinity as shown in Fig.
\ref{Hnx}. We can build up the Hilbert space by introducing the the metric
operator $\eta$ such that;%
\[
\eta H\eta^{-1}=H^{\dagger}.
\]
Or%
\[
\rho H\rho^{-1}=h,
\]
where $h$ is the equivalent Hermitian Hamiltonian. Note that;
\begin{align*}
\eta &  =\rho^{2},\\
\rho &  =\exp \left(  -\frac{1}{2}x^{2}\right)  .
\end{align*}
In using Baker--Campbell--Hausdorff formula, we get;
\[
\rho H\rho^{-1}=H+\left[  \alpha x^{2},H\right]  +\frac{1}{2}\left[  \alpha
x^{2},\left[  \alpha x^{2},H\right]  \right]  +.....
\]

\begin{align*}
\rho H\rho^{-1}  &  =p^{2}+2ixp+\left[  \alpha x^{2},p^{2}+2ixp\right]
+\frac{1}{2}\left[  \alpha x^{2},\left[  \alpha x^{2},p^{2}+2ixp\right]
\right] \\
&  =p^{2}+x^{2}-1=h.
\end{align*}

In fact, the eigen functions $H_{n}\left(  x\right)  $ do not vanish at
$x\rightarrow \infty$ but the transition amplitudes
\[
T_{ij}=\int \psi_{i}^{\ast}\eta \psi_{j}dx=\int H_{n}\left(  x\right)
\exp \left(  -x^{2}\right)  H_{m}\left(  x\right)  dx,
\]
are finite and calculable although the problem has been treated on the real
axis on which the eigen functions $H_{n}\left(  x\right)  $ violate the
condition $H_{n}\left(  x\right)  \rightarrow0 $ as $\left \vert x\right \vert
\rightarrow \infty$. This example shows us that some non-Hermitian theories can
be treated on a contour on which the eigen functions do not vanish at its
ends. Nevertheless, the structure of the metric operator can turn the
transition amplitudes finite. In this work, we will discuss some contours
which connect two adjacent Stokes wedges of the $\mathcal{PT}$-Symmetric
$(-x^{4})$ and aim to find the metric that turns the transition amplitudes
finite. To achieve our goal, we will follow the method of canonical transformations.

In quantum mechanics, canonical transformations that represent translation
and/or scaling of the position variable $x$ preserve the physical content of a
theory \cite{canonical}. So in principle, one might shift and/or scale the
real variable $x$ in the contour $z=-2i\sqrt{1+ix}$ in order to obtain another
complex contour which might connect two adjacent Stokes wedges of the theory
while the physical content of the theory stays the same. We will show that the
Hilbert spaces associated to contours connecting non-adjacent Stokes wedges
and those connecting adjacent Stokes wedges are isomorphic to each other.

To start, consider the $\mathcal{PT}$-symmetric Hamiltonian of the form;%
\begin{equation}
H=p^{2}-x^{4}. \label{orig}%
\end{equation}
The usual recipe of discussing this Hamiltonian is to replace the real
variable $x$ by a contour $z(x)$ in the complex plane \cite{Jones}. Any
contour $z\left(  x\right)  $ would represent a canonical transformation that
takes the Hamiltonian $H$ to another equivalent non-Hermitian Hamiltonian
$H_{1}$ where,%
\begin{equation}
H_{1}=\left(  \frac{\partial z\left(  x\right)  }{\partial x}\right)
^{-1}\left(  \frac{\partial z\left(  x\right)  }{\partial x}\right)
^{-1}p^{2}-i\left(  \frac{\partial z\left(  x\right)  }{\partial x}\right)
^{-1}\frac{\partial}{\partial x}\left(  \frac{\partial z\left(  x\right)
}{\partial x}\right)  ^{-1}p-\left(  z\left(  x\right)  \right)  ^{4}.
\label{H1}%
\end{equation}
Let us parametrize the contour $z\left(  x\right)  =-2i\sqrt{1+ix}$ chosen by
Jones and Mateo in Ref. \cite{Jones} such that;%
\begin{equation}
z\left(  x\right)  =a\sqrt{b+icx}. \label{paramc}%
\end{equation}
Some sets of the parameters $a$, $b$ and $c$ can define complex contours that
connect either adjacent or non-adjacent Stokes wedges. In using the
parametrized contour in Eq.(\ref{paramc}), we get the Hamiltonian in
Eq.(\ref{orig}) transformed to the non-Hermitian form;
\begin{equation}
H_{1}=-\frac{4}{a^{2}c^{2}}\left(  b+icx\right)  p^{2}-\frac{2}{a^{2}c}%
p-a^{4}\left(  b+icx\right)  ^{2}.
\end{equation}
This Hamiltonian might be $\mathcal{PT}$-symmetric or not depending on the
parameters $a,b$ and $c$. Regardless of being $\mathcal{PT}$-symmetric \ or
not, one may aim to obtain an equivalent Hermitian Hamiltonian by applying a
transformation of the form;%
\begin{equation}
\rho=\exp \left(  fp^{3}+gp\right)  , \label{similar}%
\end{equation}
where $f$ and $g$ are $C$-number parameters. The transformation $\rho$
transforms $x$ as;
\begin{align}
\rho x\rho^{-1}  &  =x+\left[  \left(  fp^{3}+gp\right)  ,x\right] \nonumber \\
&  =\left(  x-3ifp^{2}-ig\right)  ,
\end{align}
and $\rho p\rho^{-1}=p.$ If we set;
\[
g=-\frac{b}{c}\  \text{and\  \ }f=-\frac{2}{3a^{6}c^{3}},
\]
we get the following Hamiltonian;%
\begin{equation}
h=\frac{4}{a^{8}c^{4}}p^{4}+\frac{2}{a^{2}c}p+a^{4}c^{2}x^{2}, \label{Her}%
\end{equation}
which is Hermitian provided that $a^{2}c$ is real. In this case, the parameter
$f$ \ in Eq. (\ref{similar}) is also real but $g$ need not to be real. The
Hamiltonian $h$ in Eq.(\ref{Her}) with $a^{2}c$ real is equivalent to the
Hermitian Hamiltonian obtained in Ref.\cite{Jones} since one can apply the
canonical transformation $x\rightarrow \frac{2p}{a^{2}c},\ p\rightarrow
-\frac{a^{2}cx}{2}$ to obtain;%
\begin{equation}
h\rightarrow h_{1}=p^{2}+4x^{4}-2x,
\end{equation}
which is exactly the Hamiltonian obtained in Ref.\cite{Jones} ( with the
coupling $g$ there is taken here to equal $1$).

In the above discussions, there is no constraint on the parameter $b$.
However, since $\rho$ should be Hermitian, the ratio $\frac{b}{c}$ should be real.

If we choose $b=1$, $c=1$ and $a=i,$ we obtain $a^{2}c=-1$ \ and thus the
contour $z(x)$ ( the contour labeled by (2) in Fig.\ref{inst}) takes the form;%
\begin{equation}
z\left(  x\right)  =a\sqrt{b+icx}=i\sqrt{ix+1}.
\end{equation}
From Fig.\ref{inst}, this contour lies in the upper half of the complex plane
and connects two non-adjacent Stokes wedges that are $\mathcal{PT}$-symmetric
to each other. With this contour, the equivalent Hermitian Hamiltonian in
Eq.(\ref{Her}) takes the form;%
\begin{align}
h  &  =h=4p^{4}-2p+x^{2}\nonumber \\
&  \equiv p^{2}+4x^{4}-2x,
\end{align}
where in the second line we used the canonical transformation $x\rightarrow
-p,$ $p\rightarrow x$. This Hamiltonian is the same as the one obtained in
Ref. \cite{Jones} although the contour lies in the upper half. Note that, the
$WKB$ approximation, in momentum space, for the wave function $\widetilde
{\phi}\left(  p\right)  $ of the Hamiltonian in Eq. (\ref{H1}) gives;
\begin{equation}
\widetilde{\phi}\left(  p\right)  \sim \sqrt[3]{\sqrt{2\left(  2p^{3}-1\right)
}+2p^{\frac{3}{2}}}\exp \left(  \frac{-2}{3}p^{3}+p-\frac{\left \vert
p\right \vert \sqrt{2p\left(  2p^{3}-1\right)  }}{3}\right)  ,
\end{equation}
which goes to zero as $\left \vert p\right \vert \rightarrow \infty$.

Another contour that leads to the same equivalent Hamiltonian is
$z(x)=\sqrt{1+ix}$ \ (labeld by (3) in Fig. (\ref{inst})). This contour is
neither $\mathcal{PT}$-symmetric nor it connects non-adjacent Stokes wedges
and thus the two possible solutions of the Schrodinger equation can not be
finite at the two ends of the contour. This contour surprisingly results in
the same equivalent Hermitian Hamiltonian. In fact, working with this contour
is similar to investigate the Hermite differential equation on the real line
where the eigne functions blow up at infinity. However, the weight function (
metric operator) $\exp \left(  -x^{2}\right)  $ turns the probabilities
$H_{n}\left(  x\right)  \exp \left(  -\frac{1}{2}x^{2}\right)  H_{n}\left(
x\right)  $ finite everywhere on the real line. Similarly, the $\mathcal{PT}%
$-symmetric $\left(  -x^{4}\right)  $ theory on the contour $z(x)=\sqrt{1+ix}$
have eigne functions $\phi \left(  x\right)  $ that violates the condition
$\phi \left(  x\right)  \rightarrow0$ as $\left \vert x\right \vert
\rightarrow \infty$ but the probability amplitude $\phi \left(  x\right)
\eta \phi \left(  x\right)  $ is finite as $\left \vert x\right \vert
\rightarrow \infty.$ These features can be verified by considering the $WKB$
approximation of momentum space wave function $\widetilde{\phi}\left(
p\right)  $ which teaks the form;
\begin{equation}
\widetilde{\phi}\left(  p\right)  =\left(  \frac{1}{\sqrt[3]{\sqrt{2}%
p^{\frac{3}{2}}+\sqrt{2p^{3}+1}}}\exp \left(  \frac{2}{3}p^{3}+p-\frac{1}%
{3}\sqrt{4p^{6}+2p^{3}}\right)  \right)  .
\end{equation}
Note that $\widetilde{\phi}\left(  p\right)  \rightarrow \infty$ as
$p\rightarrow \infty$ and $\widetilde{\phi}\left(  p\right)  \rightarrow0$ as
$p\rightarrow-\infty$. In fact, this is expected as the contour $z(x)=\sqrt
{1+ix}$ connects two adjacent Stokes wedges that are not $\mathcal{PT}%
$-symmetric to one another. However, the probability takes the form,
\begin{equation}
\widetilde{\phi}\left(  p\right)  \eta \widetilde{\phi}\left(  p\right)
\sim \left(  \frac{1}{\sqrt[3]{\sqrt{2}p^{\frac{3}{2}}+\sqrt{2p^{3}+1}}}%
\exp \left(  -\frac{1}{3}\sqrt{4p^{6}+2p^{3}}\right)  \right)  ^{2},
\end{equation}
where $\eta=e^{-\frac{4}{3}p^{3}-2p}$ is the metric operator. Accordingly,
although the wave function $\widetilde{\phi}\left(  p\right)  $ blows up at
one end of the contour, the metric turns the theory finite the same way the
weight function does with the Hermite polynomials.

The last case we study here is the contour $\sqrt{ix}$ which also have the
same Hermitian Hamiltonian as the contour $-2i\sqrt{1+ix}$. In fact, this
contour can be considered in the lower or the upper half of the complex plane.
This is because, for each value of $x$, there exist two roots one of postie
imaginary part and the other of negative imaginary one. Either taking the
upper or the lower root will result in a wave function $\widetilde{\phi
}\left(  p\right)  $ that goes to zero as $\left \vert p\right \vert
\rightarrow \infty$. This can be easily checkd by the $WKB$ approximation which
results in;%
\begin{equation}
\widetilde{\phi}\left(  p\right)  =\frac{1}{\sqrt[3]{\sqrt{2}p^{\frac{3}{2}%
}+\sqrt{2p^{3}+1}}}\exp \left(  \frac{2}{3}p^{3}-\frac{1}{3}\sqrt{4p^{6}%
+2p^{3}}\right)  .
\end{equation}

Now, we have shown that contours from upper and lower halves in the complex
$x$ plane can lead to the same equivalent Hermitian Hamiltonian. To prove the
equivalence, one has to show that all the transition amplitudes are also the
same for all contours. To show this, one consider the metric operator $\eta$
from Eq.(\ref{similar}) which can be written as;%
\begin{equation}
\eta=\rho^{2}=\exp \left(  -\frac{4p^{3}}{3a^{6}c^{3}}-\frac{2b}{c}p\right)
.\label{metric}%
\end{equation}
Thus for different parameters (\textit{i.e} different contours), different
metric operators will define different Hilbert spaces and one may wonder if
these Hilbert spaces are equivalent (isomorphic). To discuss this point; let
$\psi_{i}$ are the eigen functions of the differential equation associated
with the Hamiltonian $H_{1}$,
\begin{equation}
\left(  \left(  \frac{\partial z_{1}\left(  x\right)  }{\partial x}\right)
^{-1}\left(  \frac{\partial z_{1}\left(  x\right)  }{\partial x}\right)
^{-1}p^{2}-i\left(  \frac{\partial z_{1}\left(  x\right)  }{\partial
x}\right)  ^{-1}\frac{\partial}{\partial x}\left(  \frac{\partial z_{1}\left(
x\right)  }{\partial x}\right)  ^{-1}p-\left(  z_{1}\left(  x\right)  \right)
^{4}\right)  \psi_{i}=E_{i}\psi_{i},
\end{equation}
for the complex contour $z_{1}\left(  x\right)  =a_{1}\sqrt{b_{1}+ic_{1}x}$
and $\phi_{i}$ are the eigen functions of the differential equation;%
\[
\left(  \left(  \frac{\partial z_{2}\left(  x\right)  }{\partial x}\right)
^{-1}\left(  \frac{\partial z_{2}\left(  x\right)  }{\partial x}\right)
^{-1}p^{2}-i\left(  \frac{\partial z_{2}\left(  x\right)  }{\partial
x}\right)  ^{-1}\frac{\partial}{\partial x}\left(  \frac{\partial z_{2}\left(
x\right)  }{\partial x}\right)  ^{-1}p-\left(  z_{2}\left(  x\right)  \right)
^{4}\right)  \phi_{i}=E_{i}\phi_{i},
\]
associated with the contour $z_{2}\left(  x\right)  =a_{2}\sqrt{b_{2}+ic_{2}%
x}.$ To obtain the map that transforms $z_{1}\left(  x\right)  $to
$z_{2}\left(  x\right)  $, one rewrites $z_{2}\left(  x\right)  $ in the
form;
\begin{equation}
z_{2}\left(  x\right)  =\sqrt{a_{2}^{2}b_{2}+ia_{2}^{2}c_{2}x}=\sqrt{a_{1}%
^{2}b_{1}+i\frac{a_{2}^{2}c_{2}}{a_{1}^{2}c_{1}}a_{1}^{2}c_{1}\left(
x-i\left(  \frac{b_{2}}{c_{2}}-\frac{a_{1}^{2}b_{1}}{a_{2}^{2}c_{2}}\right)
\right)  }.
\end{equation}
So, we can obtain $z_{2}$ from scaling $x$ in $z_{1}$ by $\frac{a_{2}^{2}%
c_{2}}{a_{1}^{2}c_{1}}$ and then shifting $x$ by $-i\left(  \frac{b_{2}}%
{c_{2}}-\frac{a_{1}^{2}b_{1}}{a_{2}^{2}c_{2}}\right)  $. These operations can
be represented by the consequent transformations; $\zeta_{1}=\exp \left(
\frac{i}{2}\ln \beta \left \{  x,p\right \}  \right)  $ and $\zeta_{2}=\exp \left(
\gamma p\right)  $ \cite{canonical} where;
\begin{equation}
\beta=\frac{a_{2}^{2}c_{2}}{a_{1}^{2}c_{1}}\text{, \ }\gamma=\frac{b_{2}%
}{c_{2}}-\frac{a_{1}^{2}b_{1}}{a_{2}^{2}c_{2}}.
\end{equation}
In other words, the map $\zeta=\zeta_{2}\zeta_{1}$ is the transformation
mapping the complex contour $z_{1}\left(  x\right)  $ to $z_{2}\left(
x\right)  $ and thus we have $\phi_{i}=\zeta \psi_{i}$. To show that $\zeta$ is
an isomorphism, we need to show that $\zeta$ preserve the inner product or in
other words to show that the transition amplitudes equality of the form;
$\langle \psi_{i}\left \vert \eta_{1}\right \vert \psi_{j}\rangle=\langle \phi
_{i}\left \vert \eta_{2}\right \vert \phi_{j}\rangle$ holds. For that, consider
the transition amplitudes associated with $z_{1}\left(  x\right)  =a_{1}%
\sqrt{b_{1}+ic_{1}x}$ to be;%
\[
\langle \psi_{i}\left \vert \eta_{1}\right \vert \psi_{j}\rangle=\langle \psi
_{i}\left \vert \exp \left(  -\frac{4}{3a_{1}^{6}c_{1}^{3}}p^{3}-\frac{2b_{1}%
}{c_{1}}p\right)  \right \vert \psi_{j}\rangle.
\]
Using the map $\zeta$ one can show that;
\begin{align}
\langle \psi_{i}\left \vert \eta_{1}\right \vert \psi_{j}\rangle &  =\langle
\zeta^{-1}\phi_{i}\left \vert \exp \left(  -\frac{4}{3a_{1}^{6}c_{1}^{3}}%
p^{3}-\frac{2b_{1}}{c_{1}}p\right)  \right \vert \zeta^{-1}\phi_{j}%
\rangle \nonumber \\
&  =\langle \zeta_{1}^{-1}\zeta_{2}^{-1}\phi_{i}\left \vert \exp \left(
-\frac{4}{3a_{1}^{6}c_{1}^{3}}p^{3}-\frac{2b_{1}}{c_{1}}p\right)  \right \vert
\zeta_{1}^{-1}\zeta_{2}^{-1}\phi_{j}\rangle \nonumber \\
&  =\langle \phi_{i}\left \vert \exp \left(  -\frac{4}{3a_{1}^{6}c_{1}^{3}%
}\left(  \frac{p}{\beta}\right)  ^{3}-\frac{2b_{1}}{c_{1}}\left(  \frac
{p}{\beta}\right)  -2\gamma p\right)  \right \vert \phi_{j}\rangle \\
&  =\langle \phi_{i}\left \vert \eta_{2}\right \vert \phi_{j}\rangle \nonumber
\end{align}
where $\eta_{2}=\exp \left(  -\frac{4}{3}\frac{p^{3}}{a_{2}^{6}c_{2}^{3}}%
-\frac{2b_{2}}{c_{2}}p\right)  $. This shows that the transformation $\zeta$
is an isomorphism that maps the Hilbert space associated with $z_{1}\left(
x\right)  $ to the Hilbert space associated with $z_{2}\left(  x\right)  $.

Let us give some examples for the contours $z_{1}$ and keep always $z_{2}$ to
be the one chosen in Ref. \cite{Jones}, $z_{2}=-2i\sqrt{1+ix}$.  Let
$z_{1}=i\sqrt{1+ix}$,  the metric operator is then given by;
\[
\eta_{1}=\exp \left(  \frac{4}{3}p^{3}-2p\right)  .
\]
The transition amplitudes are then
\[
\langle \psi_{i}\left \vert \eta_{1}\right \vert \psi_{j}\rangle=\langle \psi
_{i}\left \vert \exp \left(  \frac{4}{3}p^{3}-2p\right)  \right \vert \psi
_{j}\rangle.
\]
Based on the above discussion, the map
\[
\zeta=\exp \left(  \gamma p\right)  \exp \left(  \frac{i}{2}\ln \beta \left \{
x,p\right \}  \right)  ,
\]
transforms the contour $z_{1}=i\sqrt{1+ix}$ to the contour $z_{2}%
=-2i\sqrt{1+ix}$ where
\begin{align*}
\beta &  =\frac{a_{2}^{2}c_{2}}{a_{1}^{2}c_{1}}\text{, \ }\gamma=\frac{b_{2}%
}{c_{2}}-\frac{a_{1}^{2}b_{1}}{a_{2}^{2}c_{2}},\\
a_{2} &  =-2i\text{, }b_{2}=1\text{, \ }c_{2}=1.
\end{align*}
 So,
\begin{align}
\langle \psi_{i}\left \vert \eta_{1}\right \vert \psi_{j}\rangle &  =\langle
\psi_{i}\left \vert \exp \left(  \frac{4}{3}p^{3}-2p\right)  \right \vert
\psi_{j}\rangle \nonumber \\
&  =\langle \phi_{i}\left \vert \left(  \zeta^{-1}\right)  ^{\dagger}\exp \left(
\frac{4}{3}p^{3}-2p\right)  \zeta^{-1}\right \vert \phi_{j}\rangle \nonumber \\
&  =\langle \phi_{i}\left \vert \exp \left(  \frac{1}{48}p^{3}-2p\right)
\right \vert \phi_{j}\rangle.
\end{align}
One can realize that $\eta_{2}=\exp \left(  \frac{1}{48}p^{3}-2p\right)  $ is
exactly the metric operator obtained in Ref.\cite{Jones}.  For the contour
$z_{1}=\sqrt{1+ix}$  with the contour $z_{2}=-2i\sqrt{1+ix}$ ,  we have also
the equivalence relation;%

\begin{align}
\langle \psi_{i}\left \vert \eta_{1}\right \vert \psi_{j}\rangle &  =\langle
\psi_{i}\left \vert \exp \left(  -\frac{4}{3}p^{3}-2p\right)  \right \vert
\psi_{j}\rangle \nonumber \\
&  =\langle \phi_{i}\left \vert \exp \left(  \frac{1}{48}p^{3}-2p\right)
\right \vert \phi_{j}\rangle.
\end{align}
 For  the contour $z_{1}=\sqrt{ix}$, we get  the equivalence relation;%

\begin{align}
\langle \psi_{i}\left \vert \eta_{1}\right \vert \psi_{j}\rangle &  =\langle
\psi_{i}\left \vert \exp \left(  -\frac{4}{3}p^{3}\right)  \right \vert \psi
_{j}\rangle \nonumber \\
&  =\langle \phi_{i}\left \vert \exp \left(  \frac{1}{48}p^{3}-2p\right)
\right \vert \phi_{j}\rangle.
\end{align}

In conclusion, we showed that one can employ complex contours in the upper
half of the complex plane for the $\mathcal{PT}$-symmetric $\left(
-x^{4}\right)  $ theory. For these contours, the amplitudes stay the same as
those associated with $\mathcal{PT}$-symmetric contours in the lower half.
Besides, we showed that a complex contour that is symmetric about the real
line can preserve the physical content of the theory. For this contour, the
solutions of the Schrodinger equation cannot be finite at both ends of the
contour but the metric operator has been shown to turn the transition
amplitudes finite.

To elucidate our idea, we parametrized the contour $z_{2}\left(  x\right)
=-2i\sqrt{1+ix}$ studied in Ref. \cite{Jones}. The parametrized contour takes
the form $z\left(  x\right)  =a\sqrt{b+icx}$, where $a$, $b$ and $c$ are
$C$-number parameters. The parameters $a$, $b$ and $c$ can be varied in such a
way that results in contours connecting adjacent Stokes wedges.

For the parametrized contour $a\sqrt{b+icx}$, we found that there exists
equivalent Hermitian Hamiltonian if $a^{2}c$ is kept real. This Hermitian
Hamiltonian is found to be $b$-independent and is exactly the same as the one
associated with the contour $z_{2}\left(  x\right)  =-2i\sqrt{1+ix}$ from
Ref.\cite{Jones}.

We showed that all the Hilbert spaces associated with different parameter
choices are isomorphic to each other. The most interesting realization is that
when considering the contour $z\left(  x\right)  =\sqrt{1+ix}$ which connects
two adjacent Stokes wedges. For this contour, the $WKB$ approximation for the
wave function shows that it blows up at one end of the contour. In calculating
the transition amplitudes using this contour, we found that they are exactly
the same as those obtained by considering the contour $z\left(  x\right)
=-2i\sqrt{1+ix}$ which is $\mathcal{PT}$-symmetric as will as connecting two
non-adjacent Stokes wedges. The point is that while the wave function
associated with the contour $\sqrt{1+ix}$ \ violates the condition
$\phi \left(  x\right)  \rightarrow0$ as $\left \vert x\right \vert
\rightarrow \infty$, the metric turns the probability amplitude of the form
$\phi^{\ast}\left(  x\right)  \eta \phi \left(  x\right)  $ finite everywhere
and thus the theory is square integrable.

\newpage

\end{document}